\documentclass[preprint,aps,12pt,showpacs,nofootinbib,tightenlines]{revtex4}
\usepackage{amsmath}
\usepackage{amssymb}
\usepackage{epsfig}
\usepackage{graphicx}
\begin{document}
\preprint{NJNU-TH-07-03}

\newcommand{\beq}{\begin{eqnarray}}
\newcommand{\eeq}{\end{eqnarray}}
\newcommand{\non}{\nonumber\\ }

\newcommand{\acp}{{\cal A}_{CP}}
\newcommand{\etap}{\eta^{(\prime)} }
\newcommand{\etapr}{\eta^\prime }
\newcommand{\pb}{\phi_B}
\newcommand{\pp}{\phi_{\pi}}
\newcommand{\pe}{\phi_{\eta}}
\newcommand{\pepr}{\phi_{\etap}}
\newcommand{\ppp}{\phi_{\pi}^P}
\newcommand{\pep}{\phi_{\eta}^P}
\newcommand{\peprp}{\phi_{\etap}^P}
\newcommand{\ppt}{\phi_{\pi}^t}
\newcommand{\pet}{\phi_{\eta}^t}
\newcommand{\peprt}{\phi_{\etap}^t}
\newcommand{\fb}{f_B }
\newcommand{\fpi}{f_{\pi} }
\newcommand{\feta}{f_{\eta} }
\newcommand{\fetap}{f_{\etap} }
\newcommand{\rpi}{r_{\pi} }
\newcommand{\re}{r_{\eta} }
\newcommand{\rep}{r_{\etap} }
\newcommand{\mb}{m_B }
\newcommand{\mop}{m_{0\pi} }
\newcommand{\moe}{m_{0\eta} }
\newcommand{\moep}{m_{0\etap} }

\newcommand{\psl}{ p \hspace{-1.8truemm}/ }
\newcommand{\nsl}{ n \hspace{-2.2truemm}/ }
\newcommand{\vsl}{ v \hspace{-2.2truemm}/ }
\newcommand{\epsl}{\epsilon \hspace{-1.8truemm}/\,  }

\def \epjc{ Eur.Phys.J. C }
\def \jpg{  J. Phys. G }
\def \npb{  Nucl. Phys. B }
\def \plb{  Phys. Lett. B }
\def \pr{  Phys. Rep. }
\def \prd{  Phys. Rev. D }
\def \prl{  Phys. Rev. Lett.  }
\def \zpc{  Z. Phys. C  }
\def \jhep{ J. High Energy Phys.  }

\title{$B^0 \to \omega\eta^{(\prime)}$ and  $\phi\eta^{(\prime)}$ decays
 in the perturbative QCD approach}
\author{Dong-qin Guo} \email{medongqin@163.com}
\author{Xin-fen Chen} \email{chenxinfen@163.com}
\author{Zhen-jun Xiao} \email{xiaozhenjun@njnu.edu.cn}
\affiliation{Department of Physics and Institute of Theoretical Physics, Nanjing Normal
University, Nanjing, Jiangsu 210097, P.R.China}
\date{\today}
\begin{abstract}
We calculate the branching ratios and CP-violating asymmetries for
$B^0 \to\omega \eta $, $\omega \eta^\prime$, $\phi \eta$ and $\phi
\eta^\prime$ decays in the perturbative QCD (pQCD) factorization
approach. The pQCD predictions for the CP-averaged branching ratios
are
 $Br(B^0 \to \omega\eta) = \left ( 2.7 ^{+1.1}_{-1.0}\right )\times 10^{-7}$,
 $Br(B^0 \to \omega \eta^\prime) = \left (0.75 ^{+0.37}_{-0.33}\right ) \times 10^{-7}$,
 and
 $Br(B^0 \to \phi\eta) = \left ( 6.3 ^{+3.3}_{-1.9}\right ) \times 10^{-9}$,
 $Br(B^0 \to \phi\eta^{\prime}) = \left ( 7.3^{+3.5}_{-2.6}\right ) \times 10^{-9}$
which are consistent with currently available  experimental upper limits.
The inclusion of the gluonic contribution can change the branching ratios of
$B \to \omega(\phi) \eta^\prime$ decays by about $10\%$.
The direct CP-violating asymmetries for $B^0 \to \omega \eta$ and $\omega
\eta^\prime$ decays are generally large in size.
\end{abstract}


\maketitle

\section{Introduction}

As is well-known, the two-body charmless B meson decays provide a good place for
testing the standard model (SM) and  searching for the new physics signals.
Among various $B \to M_1 M_2$ ( here $M_i$ refers to the light pseudo-scalar or
vector mesons ) decay channels, the decays involving the $\eta$ or $\etapr$ meson in the final state  have
been studied extensively during the past decade because of the
so-called $K\eta^\prime$ puzzle or other special features.

In Ref.~\cite{bn03b}, for example, many decay modes involving $\etap$ meson
were studied in the QCD factorization (QCDF) approach \cite{bbns99}.
In the pQCD factorization approach \cite{pqcd}, on the other hand,
the $B \to K \etap$, $\rho \etap$, $\pi \etap$ and $\etap\etap$ decays have been calculated in
Refs.~\cite{ekou2,liu06,wang06,guo06a}. The $B_{s} \to (\pi, \rho, \omega, \phi )\eta^{(\prime)}$
decays, furthermore,  have also been studied  in the pQCD approach \cite{bsdecay}.

In this paper, we will perform the leading order pQCD calculation for $B \to \omega
\etap$ and $\phi \etap$ decays. Besides the usual factorizable contributions, we here are able to
evaluate the non-factorizable and the annihilation contributions as well.
Besides the dominant contributions from the $q\bar{q}$ components of $\etap$ mesons,
the contribution from the possible gluonic component of $\etap$ meson will also be included
by using the formulae as given in Ref.~\cite{li0609}.

On the experimental side, only the upper limits on the branching ratios of $B \to \omega(\phi) \etap$ decays
are available now \cite{hfag06,0701046ex}
\beq
Br(B^0 \to\phi\eta)< 0.6\times 10^{-6}, \quad
Br(B^0 \to \phi \eta^\prime)<1.0\times 10^{-6}. \label{eq:exp1} \\
Br(B^0 \to \omega\eta)< 1.9 \times 10^{-6}, \quad
Br(B^0 \to \omega\eta^\prime)< 2.8\times 10^{-6}. \label{eq:ulimits}
\eeq
But these decays could be measured with good precision in the forthcoming LHC-b experiments if their
decay rates are larger than $10^{-8}$.

For $B \to \omega (\phi) \etap$ decays considered here, it is generally believed that the $q\bar{q}$
component of $\etap$ meson provide the dominant contribution, but it is  still very difficult
to calculate reliably the gluonic contribution from the possible glunoic component of $\etap$ meson
\cite{ali03a,li0609}.
Of course, great efforts have been made for this problem and some progress have been achieved recently
\cite{ali03a,li0609} in order to explain the large $B \to K \eta^\prime$ decay rates.
In Ref.~\cite{li0609}, for instance, the authors found that the possible
gluonic contribution is small for both $B \to \eta$ and $B \to \eta^\prime$ form factors
\cite{li0609,li0612}.

This paper is organized as follows. In Sec.~\ref{sec:f-work}, we
give a brief review for the PQCD factorization approach. In
Sec.~\ref{sec:p-c}, we calculate analytically the related Feynman
diagrams and present the various decay amplitudes for the studied
decay modes. In Sec.~\ref{sec:n-d}, we show the numerical results
for the branching ratios and CP asymmetries of $B \to   \omega(\phi) \eta^{(\prime)}$
decays. The summary and some discussions are included in the final section.

\section{ the theoretical framework }\label{sec:f-work}

At present, there exist two popular factorization approaches to calculate
the  hadronic matrix element $<M_1 M_2|O_i|B>$: the QCDF approach \cite{bbns99}
and the pQCD approach \cite{pqcd,cl97,lb80}.
The pQCD approach has been developed earlier from the QCD
hard-scattering approach \cite{lb80}. Some elements of this
approach are also present in the QCD factorization approach \cite{bbns99,bn03b}.
The two major differences between these two approaches are
(a) the form factors are calculable perturbatively
in pQCD approach, but taken as the input parameters extracted from
other experimental measurements in the QCDF approach; and (b) the
annihilation contributions are calculable and play an important
role in producing CP violation for the considered decay modes in
pQCD approach, but it could not be evaluated reliably in QCDF
approach. Of course, the assumptions behind the pQCD approach, specifically the possibility to
calculate the form factors perturbatively, are still under
discussion~\cite{ds02}. More efforts are needed to clarify these problems.

In pQCD approach, the decay amplitude is separated into soft ($\Phi$), hard(H), and harder(C) dynamics
characterized by different energy scales $(t,m_b,M_W)$. It is conceptually written as the convolution,
\beq
{\cal A}(B \to M_1 M_2)\sim \int\!\! d^4k_1 d^4k_2 d^4k_3\
\mathrm{Tr} \left [ C(t) \Phi_B(k_1) \Phi_{M_1}(k_2)
\Phi_{M_2}(k_3) H(k_1,k_2,k_3, t) \right ], \label{eq:con1}
\eeq
where the $k_i$ are the momenta of the light quarks included in each of the mesons,
and $\mathrm{Tr}$ denotes the trace over Dirac and color indices.
$C(t)$ is the Wilson coefficient which results from the radiative
corrections at short distance. In the above convolution, $C(t)$
includes the harder dynamics at larger scale than $M_B$ scale and
describes the evolution of local $4$-Fermi operators from $m_W$ (the
$W$ boson mass) down to $t\sim\mathcal{O}(\sqrt{\bar{\Lambda} M_B})$
scale, where $\bar{\Lambda}\equiv M_B -m_b$. The function
$H(k_1,k_2,k_3,t)$ is the hard part and can be  calculated
perturbatively. The function $\Phi_M$ is the wave function which
describes hadronization of the quark and anti-quark to the meson
$M$. While the function $H$ depends on the process considered, the
wave function $\Phi_M$ is independent of the specific decay process. Using
the wave functions determined from other well measured processes,
one can make quantitative predictions here.

Since the b quark is rather heavy, we consider the $B$ meson at rest
for simplicity. It is convenient to use light-cone coordinate $(p^+, p^-, {\bf p}_T)$ to
describe the meson's momenta,
\beq
p^\pm =\frac{1}{\sqrt{2}} (p^0 \pm p^3), \quad and \quad {\bf p}_T = (p^1, p^2).
\eeq
Using these coordinates the $B$ meson and the two final
state meson momenta can be written as
\beq
P_1 =\frac{M_B}{\sqrt{2}} (1,1,{\bf 0}_T), \quad
P_2 =\frac{M_B}{\sqrt{2}} (1,r_{\omega(\phi)}^{2},{\bf 0}_T), \quad
P_3 =\frac{M_B}{\sqrt{2}} (0,1-r_{\omega(\phi)}^{2},{\bf 0}_T),
\eeq
respectively, where $r^2_{\omega(\phi)}=m^2_{\omega(\phi)}/m^2_B$; and the terms proportional
to the ratio $m_{\etap}^2/m_B^2$ have been neglected.

For the $B \to \omega \eta$ decay considered here, only the $\omega$
meson's longitudinal part contributes to the decay, its polar vector
is $\epsilon_L=\frac{M_B}{\sqrt{2}M_\omega} (1,-r_\omega^2,\bf{0_T})$.
Putting the light (anti-) quark momenta in the $B$, $\omega$ and $\eta$
meson $k_1$, $k_2$, and $k_3$, respectively, we can choose
\beq
k_1 = (x_1 P_1^+,0,{\bf k}_{1T}), \quad k_2 = (x_2 P_2^+,0,{\bf
k}_{2T}), \quad k_3 = (0, x_3 P_3^-,{\bf k}_{3T}).
\eeq
Then, the integration over $k_1^-$, $k_2^-$, and $k_3^+$ in Eq.(\ref{eq:con1})
will lead to
\beq {\cal A}(B \to \omega\eta) &\sim &\int\!\! d x_1
d x_2 d x_3 b_1 d b_1 b_2 d b_2 b_3 d b_3 \non &&
\cdot \mathrm{Tr}
\left [ C(t) \Phi_B(x_1,b_1) \Phi_{\omega}(x_2,b_2) \Phi_{\eta}(x_3,
b_3) H(x_i, b_i, t) S_t(x_i)\, e^{-S(t)} \right ], \label{eq:a2}
\eeq
where $b_i$ is the conjugate space coordinate of $k_{iT}$, and
$t$ is the largest energy scale in the function $H(x_i,b_i,t)$. The
large logarithms $\ln (m_W/t)$ are included in the Wilson
coefficients $C(t)$. The large double logarithms ($\ln^2 x_i$) on
the longitudinal direction are summed by the threshold resummation, and they lead to
the function $S_t(x_i)$ which smears the end-point singularities on $x_i$.
The last term, $e^{-S(t)}$, is the Sudakov
form factor which suppresses the soft dynamics effectively
~\cite{soft}. Thus it makes the perturbative calculation of the hard
part $H$ applicable at an intermediate scale, i.e., the $M_B$ scale. We
will calculate analytically the function $H(x_i,b_i,t)$ for the
considered decays in the first order in an $\alpha_s$ expansion and
give the convoluted amplitudes in the next section.

\subsection{ Wilson Coefficients}\label{ssec:w-c}

For the $B^0 \to \omega\eta^{(\prime)}$, $B^0 \to
\phi\eta^{(\prime)}$ decays, the related weak effective Hamiltonian
$H_{eff}$ can be written as \cite{buras96}
\beq
\label{eq:heff}
{\cal H}_{eff} = \frac{G_{F}} {\sqrt{2}} \, \left[ V_{ub} V_{ud}^*
\left (C_1(\mu) O_1^u(\mu) + C_2(\mu) O_2^u(\mu) \right) - V_{tb}
V_{td}^* \, \sum_{i=3}^{10} C_{i}(\mu) \,O_i(\mu) \right] \; ,
\eeq
where $C_i(\mu)$ are Wilson coefficients at the renormalization
scale $\mu$ and $O_i$ are the four-fermion operators. For $b \to d$ transition, for example, these
operators can be written as
\beq
\begin{array}{llllll}
O_1^{u} & = &  \bar d_\alpha\gamma^\mu L u_\beta\cdot \bar u_\beta\gamma_\mu L b_\alpha\ ,
&O_2^{u} & = &\bar d_\alpha\gamma^\mu L u_\alpha\cdot \bar
u_\beta\gamma_\mu L b_\beta\ , \\
O_3 & = & \bar d_\alpha\gamma^\mu L b_\alpha\cdot \sum_{q'}\bar
 q_\beta'\gamma_\mu L q_\beta'\ ,   &
O_4 & = & \bar d_\alpha\gamma^\mu L b_\beta\cdot \sum_{q'}\bar
q_\beta'\gamma_\mu L q_\alpha'\ , \\
O_5 & = & \bar d_\alpha\gamma^\mu L b_\alpha\cdot \sum_{q'}\bar
q_\beta'\gamma_\mu R q_\beta'\ ,   & O_6 & = & \bar
d_\alpha\gamma^\mu L b_\beta\cdot \sum_{q'}\bar
q_\beta'\gamma_\mu R q_\alpha'\ , \\
O_7 & = & \frac{3}{2}\bar d_\alpha\gamma^\mu L b_\alpha\cdot
\sum_{q'}e_{q'}\bar q_\beta'\gamma_\mu R q_\beta'\ ,   & O_8 & = &
\frac{3}{2}\bar d_\alpha\gamma^\mu L b_\beta\cdot
\sum_{q'}e_{q'}\bar q_\beta'\gamma_\mu R q_\alpha'\ , \\
O_9 & = & \frac{3}{2}\bar d_\alpha\gamma^\mu L b_\alpha\cdot
\sum_{q'}e_{q'}\bar q_\beta'\gamma_\mu L q_\beta'\ ,   & O_{10} &
= & \frac{3}{2}\bar d_\alpha\gamma^\mu L b_\beta\cdot
\sum_{q'}e_{q'}\bar q_\beta'\gamma_\mu L q_\alpha'\ ,
\label{eq:operators}
\end{array}
\eeq
where $\alpha$ and $\beta$ are the $SU(3)$ color indices; $L$
and $R$ are the left- and right-handed projection operators with
$L=(1 - \gamma_5)$, $R= (1 + \gamma_5)$. The sum over $q'$ runs over
the quark fields that are active at the scale $\mu=O(m_b)$, i.e.,
$q'\epsilon\{u,d,s,c,b\}$. For the Wilson coefficients $C_i(\mu)$
($i=1,\ldots,10$), we will use the leading order (LO)
expressions, although the next-to-leading order (NLO)  results
already exist in the literature ~\cite{buras96}. This is the
consistent way to cancel the explicit $\mu$ dependence in the
theoretical formulae. For the renormalization group evolution of the Wilson coefficients
from higher scale to lower scale, we use the formulae as given in
Ref.~\cite{luy01} directly. At the high $m_W$ scale, the leading
order Wilson coefficients $C_i(M_W)$ are simple and can be found
easily in Ref.~\cite{buras96}. In the pQCD approach, the scale $t$ may be
larger or smaller than the $m_b$ scale. For the case of $ m_b < t<
m_W$, we evaluate the Wilson coefficients at $t$ scale using the leading
logarithm running equations, as given in Eq.(C1) of Ref.~\cite{luy01}.
For the case of $t < m_b$, we then evaluate the Wilson
coefficients at $t$ scale by using $C_i(m_b)$ as input
and the formulae given in Appendix D of Ref.~\cite{luy01}.

\subsection{$\phi-\omega$ mixing and  $\eta-\eta^\prime$ mixing }

For the physical isoscalars $\phi$ and $\omega$ meson, we use the
``ideal" mixing scheme \cite{pdg06}:
\beq
\phi(1020) = -s\bar{s}, \qquad \omega=\frac{1}{\sqrt{2}}\left [u\bar{u} + d\bar{d} \right ].
\label{eq:mixing1}
\eeq
Such ideal mixing are supported by the known experimental measurements \cite{pdg06}.

For the $\eta-\eta^\prime$ system, there exist two popular mixing basis: the octet-singlet basis
and the quark-flavor basis \cite{fk98,0501072}.
Here we  use  the quark-flavor basis \cite{fk98} and define
\beq
\eta_q=(u\bar{u} + d\bar{d})/\sqrt{2}, \qquad \eta_s=s\bar{s}.
\label{eq:qfbasis}
\eeq
The physical states $\eta$ and $\eta^\prime$ are related to $\eta_q$ and $\eta_s$ through a single mixing
angle $\phi$,
\beq
\left( \begin{array}{c}
\eta\\ \eta^\prime \\ \end{array} \right )
&=& U(\phi) \left( \begin{array}{c}
 \eta_q\\ \eta_s \\ \end{array} \right ) =
  \left( \begin{array}{cc}
 \cos{\phi} & -\sin{\phi} \\
 \sin{\phi} & \cos\phi \end{array} \right )
 \left( \begin{array}{c}  \eta_q\\ \eta_s \\ \end{array} \right ).
 \label{eq:e-ep}
\eeq
The corresponding decay constants $f_q, f_s, f_\eta^{q,s}$ and $f_{\eta^\prime}^{q,s}$
have been defined in Ref.~\cite{fk98} as
\beq
<0|\bar q\gamma^\mu\gamma_5 q|\eta_q(P)>  &=& -\frac{i}{\sqrt2}\,f_q\,P^\mu ,\non
<0|\bar s\gamma^\mu\gamma_5 s|\eta_s(P)>  &=& -i f_s\,P^\mu \;, \label{eq:fqfs}\\
<0|\bar q\gamma^\mu\gamma_5 q|\eta^{(\prime)}(P) >  &=& -\frac{i}{\sqrt2}\,f_{\eta^{(\prime)}}^q\,P^\mu \;,\non
<0|\bar s\gamma^\mu\gamma_5 s|\eta^{(\prime)}(P) >  &=& -i f_{\eta^{(\prime)}}^s\,P^\mu \;,
\label{eq:f11}
\eeq
while the decay constants $f_\eta^{q,s}$ and $f_{\eta^\prime}^{q,s}$ are related to $f_q$ and $f_s$
via the same mixing matrix,
\beq
\left(\begin{array}{cc}
f_\eta^q & f_\eta^s \\
   f_{\eta'}^q & f_{\eta'}^s \\
\end{array} \right)= U(\phi)\left(\begin{array}{cc}
  f_q & 0 \\   0 & f_s \\ \end{array} \right)\;.
\label{eq:f12}
\eeq
The three input parameters $f_q$, $f_s$ and $\phi$ in the quark-flavor basis
have been extracted from various related experiments \cite{fk98,0501072}
\beq
f_q = (1.07\pm 0.02) f_\pi, \quad f_s = (1.34 \pm 0.06) f_\pi, \quad \phi = 39.3^\circ \pm 1.0^\circ,
\eeq
where $f_\pi=130$ MeV. In the numerical calculations, we will use these mixing parameters
as inputs.

Although $\eta$ and $\eta^\prime$ are generally considered as a linear combination of light quark pairs
$u\bar{u}, d\bar{d}$ and $s\bar{s}$, it should be noted that a
gluonic content of $\eta^\prime$ meson may be needed to interpret the anomalously large
branching ratios of $B\to K \eta^\prime$ and $J/\Psi \to \eta^\prime \gamma$ \cite{ekou2,ekou1}.
For $B \to (\rho, \pi) \etap$ decays, however, the good agreement between the pQCD predictions
\cite{liu06,wang06}
and the measured values for their  branching ratios suggest that the gluonic
contribution of $\etap$ meson should be small.
In Ref.~\cite{li0609}, very recently, the authors calculated the flavor-singlet
contribution to the $B \to \etap$ transition form factors from the
gluonic content of the $\etap$ meson. They found that the
gluonic contribution is small for both $B \to \eta$ and $B \to \eta^\prime$ form factors.

Of course, more studies are needed to provide
a reliable and precision calculation for the gluonic contribution to
the final state involving $\etap$ meson. For the sake of
simplicity, we here firstly  neglect the possible gluonic content of
$\eta$ and $\eta^\prime$ meson, and use the quark-flavor mixing scheme for $\eta$
and $\eta^\prime$ meson as defined in Eq.~(\ref{eq:e-ep}). We will
calculate the effects of a non-zero gluonic admixture of $\eta^\prime$ in next section,
and treat them as one kind of theoretical uncertainties.

\subsection{Wave Functions}\label{ssec:w-f}

Now we present the wave functions to be used in the integration.
For the wave function of the heavy $B$ meson, we take
\beq
\Phi_{B}= \frac{1}{\sqrt{2N_c}} \left [( \psl_{B} + M_{B}) \gamma_5 \phi_{B} ({\bf k_1}) \right ].
\label{bmeson}
\eeq
Here only the contribution of the Lorentz structure $\phi_{B} ({\bf k_1})$ is taken into account, since
the contribution of the second Lorentz structure $\bar{\phi}_{B}$ is numerically small
\cite{kurimoto,luyang} and has been neglected.
For the distribution amplitude $\phi_{B}$ in Eq.~(\ref{bmeson}), we adopt the model
\beq
\phi_B(x,b) &=& N_B x^2(1-x)^2 \mathrm{exp} \left
 [ -\frac{M_B^2\ x^2}{2 \omega_{b}^2} -\frac{1}{2} (\omega_{b} b)^2\right],
 \label{eq:phib}
\eeq
where $\omega_{b}$ is a free parameter and we take
$\omega_{b}=0.4\pm 0.04$ GeV in numerical calculations, and
$N_B=91.745$ is the normalization factor for $\omega_{b}=0.4$. This
is the same wave functions as being used in
Refs.~\cite{luy01,kurimoto}, which is a best fit for most of the measured hadronic B decays.

The wave function for $d\bar{d}$ components of $\eta^{(\prime)}$ meson is  given by \cite{ekou2}
\beq
\Phi_{\eta_{d\bar{d}}}(p,x,\zeta)\equiv
\frac{i \gamma_5}{\sqrt{2N_c}} \left [ \psl \phi_{\eta_{d\bar{d}}}^{A}(x)+m_0^{\eta_{d\bar{d}}}
\phi_{\eta_{d\bar{d}}}^{P}(x)+\zeta m_0^{\eta_{d\bar{d}}} (
\vsl \nsl - v\cdot n)\phi_{\eta_{d\bar{d}}}^{T}(x) \right ],
\label{eq:ddbar}
\eeq
where $p$ and $x$ are the momentum and the momentum fraction of
$\eta_{d\bar{d}}$ respectively, while $\phi_{\eta_{d\bar{d}}}^A$,
$\phi_{\eta_{d\bar{d}}}^P$ and $\phi_{\eta_{d\bar{d}}}^T$ represent
the axial vector, pseudoscalar and tensor components of the wave
function respectively.  Following Ref.~\cite{ekou2}, we here also assume that the wave
function of $\eta_{d\bar{d}}$ is same as the $\pi$ wave function based on SU(3) flavor symmetry.
The parameter $\zeta$ is either $+1$ or $-1$ depending on the assignment of the momentum fraction $x$.
We also assume that the wave function of the $u\bar{u}$ component is the same as the wave
function of $d\bar{d}$ based on the isospin symmetry between the up and down quark.
For the wave function of the $s\bar{s}$ component, we also use the
same form as given in Eq.~(\ref{eq:ddbar}) for $d\bar{d}$ component
but with different chiral enhancement and some other changes to be
specified later.

The explicit expressions of chiral enhancement scales $m_0^{q}=m_0^{\eta_{d\bar{d}}} =m_0^{\eta_{u\bar{u}}} $
and $m_0^{s}=m_0^{\eta_{s\bar{s}}}$ are given by \cite{li0609}
\beq
m_0^q\equiv \frac {m_{qq}^2}{2m_q}&=&
\frac{1}{2m_q}[m_{\eta}^2\cos^2{\phi}+m_{\eta^\prime}^2\sin^2{\phi}
-\frac{\sqrt{2}f_s}{f_q}(m_{\eta^\prime}^2-m_{\eta}^2)\cos{\phi}sin{\phi}], \label{eq:m0q}\\
m_0^s\equiv \frac {m_{ss}^2}{2m_s}&=&
\frac{1}{2m_s}[m_{\eta^\prime}^2\cos^2{\phi}+m_{\eta}^2\sin^2{\phi}
-\frac{f_q}{\sqrt{2}f_s}(m_{\eta^\prime}^2-m_{\eta}^2)\cos{\phi}sin{\phi}],
\label{eq:m0s}
\eeq
and numerically
\beq
m_0^q= 1.07 {\rm MeV}, \qquad m_0^s=1.92 {\rm GeV}
\eeq
for $m_\eta=547.5$ MeV, $m_{\eta^\prime}=957.8$ MeV, $f_q=1.07f_\pi$, $f_s=1.34 f_\pi$ and $\phi=39.3^\circ$.

For the distribution amplitude $\phi_{\eta_{q}}^A$, $\phi_{\eta_{q}}^P$ and $\phi_{\eta_{q}}^T$,
we utilize the results for $\pi$ meson obtained from the light-cone sum rule~\cite{ball} including twist-3
contributions:
\beq
\phi_{\eta_{q(s)}}^A(x)&=&\frac{3}{\sqrt{2N_c}}f_{q(s)}x(1-x)
\left\{ 1+a_2^{\eta_{q(s)}}\frac{3}{2}\left [5(1-2x)^2-1 \right ]\right. \non
&&\left.
+ a_4^{\eta_{q(s)}}\frac{15}{8} \left [21(1-2x)^4-14(1-2x)^2+1 \right ]\right \},  \\
\phi^P_{\eta_{q(s)}}(x)&=&\frac{1}{2\sqrt{2N_c}}f_{q(s)} \left \{
1+ \frac{1}{2}\left (30\eta_3-\frac{5}{2}\rho^2_{\eta_{q(s)}}
\right ) \left [ 3(1-2x)^2-1 \right] \right.  \non
&& \left. +
\frac{1}{8}\left ( -3\eta_3\omega_3-\frac{27}{20}\rho^2_{\eta_{q(s)}}-
\frac{81}{10}\rho^2_{\eta_{q(s)}}a_2^{\eta_{q(s)}} \right )\right.\non
&& \left. \cdot  \left [ 35 (1-2x)^4-30(1-2x)^2+3 \right ] \right\} ,  \\
\phi^T_{\eta_{q(s)}}(x) &=&\frac{3}{\sqrt{2N_c}}f_{q(s)} (1-2x) \non
 && \cdot \left [ \frac{1}{6}+(5\eta_3-\frac{1}{2}\eta_3\omega_3-
\frac{7}{20}\rho_{\eta_{q(s)}}^2
-\frac{3}{5}\rho^2_{\eta_{q(s)}}a_2^{\eta_{q,s}})(10x^2-10x+1)\right ],
\eeq
with the updated Gegenbauer moments \cite{sum2,ball06}
\beq
a^{\eta_{q(s)}}_2&=&0.115, \quad  a^{\eta_{q(s)}}_4 =-0.015, \quad  \rho_{\eta_q}=2m_{q}/m_{qq};
  \non
\rho_{\eta_{s}}&=&2m_{s}/m_{ss}, \quad \eta_3=0.015, \quad \omega_3=-3.0.
 \eeq

For $B \to \omega \etap$ and $\phi \etap$ decays, only the longitudinal polarized component
$\Phi_V^L$ for $V=(\omega, \phi)$ will contribute. The wave function $\Phi_\omega^L$, for example,
can be written as
\beq
\Phi_\omega^L=\frac{1}{\sqrt{2N_c}} \left [ m_\omega \not\!\epsilon_{L}\phi_\omega(x)
+\not\!\epsilon_{L}\not\! p \phi_\omega^t(x) + m_\omega I \phi^s_\omega(x) \right ],
\label{eq:phivl}
\eeq
where the $\epsilon_L$ and ${\bf p}$ is the polarization vector and the momentum of the $\omega$ meson,
the first term in above
equation is the leading twist wave function (twist-2), while the second
and third terms are subleading twist (twist-3) wave functions.
For the case of $V=\phi$, its  wave function is the same in structure as
that defined in Eq.~(\ref{eq:phivl}), but with different mass and distribution amplitudes.

For the light meson wave function, we neglect the $b$ dependent part,
which is not important in numerical analysis. The distribution amplitudes $\phi_\phi(x)$ and
$\phi_\phi^{t,s}(x)$ are given by ~\cite{ball2}
\beq
\phi_\phi(x) &=& \frac{3}{\sqrt{6} }f_\phi  x (1-x)  ,\\
\phi_\phi^t(x) &=&  \frac{f_\phi^T }{2\sqrt{6} }
  \left\{  3 (1-2 x)^2 +1.68 C_4^{1/2} (1-2x)
  +0.69 \left[1+ (1-2 x)\ln\frac{x}{1-x}  \right] \right\},\\
\phi_\phi^s(x) &=&  \frac{3}{4\sqrt{6} } f_\phi^T
\left[3(1-2 x)(4.5-11.2x+11.2x^{2})+ 1.38\ln\frac{x}{1-x} \right] .
\eeq

For the $\omega$ meson wave function, we use  ~\cite{lu06,liying06}
\beq
\phi_\omega(x) &=& \frac{3}{\sqrt{6} } f_\omega  x (1-x)  \left[1+ 0.18C_2^{3/2} (1-2x) \right],\\
\phi_\omega^t(x) &=&  \frac{f_\omega^T }{2\sqrt{6} }
  \left\{  3 (1-2 x)^2 +0.3(1-2 x)^2  \left[5(1-2 x)^2-3  \right]   \right.  \non
 &&~~\left. +0.21 [3- 30 (1-2 x)^2 +35 (1-2 x)^4] \right\},\\
\phi_\omega^s(x) &=&  \frac{3}{2\sqrt{6} }
 f_\omega^T   (1-2x)  \left[1+ 0.76 (10 x^2 -10 x +1) \right] .
\eeq

 The relevant Gegenbauer polynomials are defined by \cite{ball2}
 \beq
 C_2^{3/2} (t) &=& \frac{3}{2} \left (5t^2-1 \right ),\\
 C_4^{1/2} (t) &=& \frac{1}{8} \left (35t^4-30t^2+3 \right ).
 \eeq

\section{Perturbative Calculations}\label{sec:p-c}

In this section, we will calculate and show the decay amplitude for
each diagram including wave functions. The hard part $H(t)$ involves
the four quark operators and the necessary hard gluon connecting the
four quark operator and the spectator quark.  We first consider $B
\to \omega \eta $ decay, and then extend the calculation to other decay modes.
Analogous to the $B \to \rho \etap$ decays in \cite{liu06}, there are also eight
type diagrams contributing to $B \to \omega \eta $ and $\omega \eta^\prime$
decay, as illustrated in Figure 1.

For $B \to \omega \eta$ decay, we firstly consider the usual factorizable diagrams 1(a) and 1(b).
The operators $O_{1,2}$ and $O_{3,4,9,10}$ are $(V-A)(V-A)$ currents;
the sum of their amplitudes is given by
\beq
F_{e}&=& 4\sqrt{2}\pi G_FC_F m_B^4\int_0^1 d x_{1} dx_{3}\,
\int_{0}^{\infty} b_1 db_1 b_3 db_3\, \phi_B(x_1,b_1) \non & &
\times \left\{ \left[(1+x_3) \pe^A(x_3, b_3) +(1-2x_3) \re
(\phi_\eta^P(x_3,b_3) +\phi_\eta^T(x_3,b_3))\right] \right. \non &&
\left.\quad  \cdot \alpha_s(t_e^1)\,
h_e(x_1,x_3,b_1,b_3)\exp[-S_{ab}(t_e^1)] \right.\non && \left. +2\re
\phi_\eta^P (x_3, b_3)
\alpha_s(t_e^2)h_e(x_3,x_1,b_3,b_1)\exp[-S_{ab}(t_e^2)] \right\}.
\label{eq:ab1}
\eeq
where the ratio $\re=m_0^\eta/m_B$ with $m_0^\eta=m_0^q$ or $m_0^s$ as defined in
Eqs.~(\ref{eq:m0q},\ref{eq:m0s});
the color factor $C_F=4/3$, $\phi_B$ and $\phi_\eta^{A,P,T}$ are the
light-cone distribution amplitudes (LCDAs) of the heavy B meson and the light $\eta$ meson.
The functions $h_e$, the scales $t_e^i$ and the Sudakov factors $S_{ab}$ will be given explicitly
in Appendix \ref{sec:app1}.

\begin{figure}[tb]
\vspace{-2 cm} \centerline{\epsfxsize=21 cm \epsffile{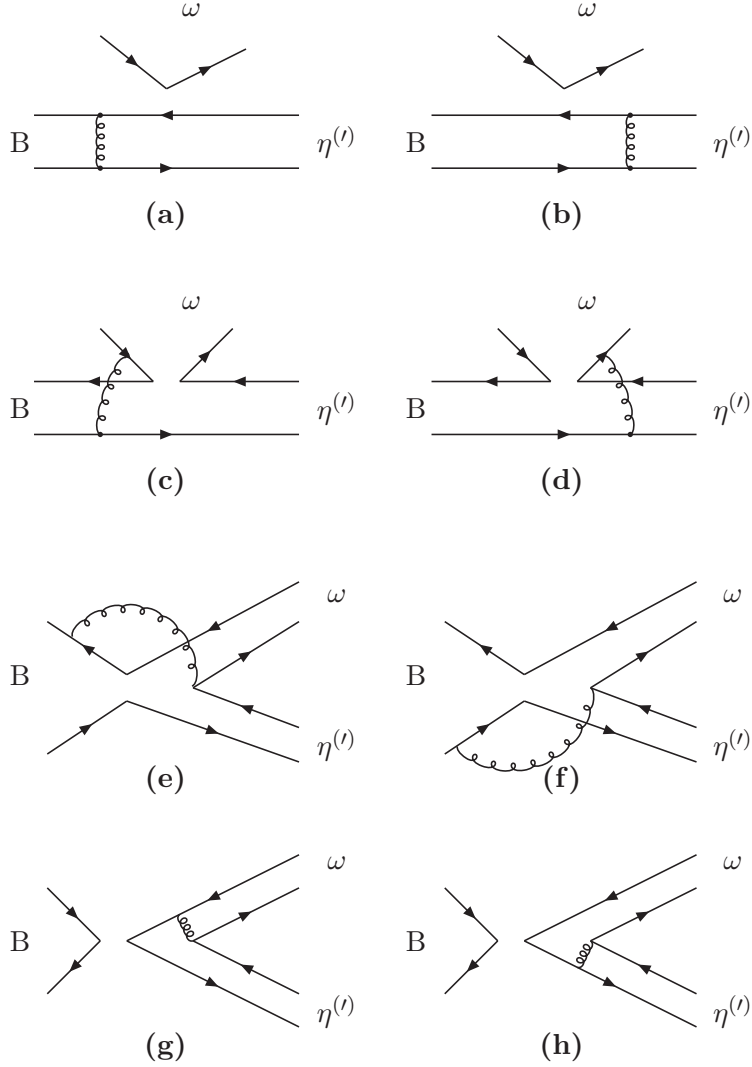}}
\vspace{-14cm}
\caption{ Feynman diagrams contributing to the $B\to\omega \etap$
 decays when the $\etap$ meson picks up the spectator quark, where
 diagram (a) and (b) contribute to the $B\to \etap$ form
 factor $F_{0,1}^{B\to \etap}$.}
 \label{fig:fig1}
\end{figure}

The operators $O_{5,6,7,8}$ have the structure of $(V-A)(V+A)$. In some decay channels, some
of these operators
contribute to the decay amplitude in a factorizable way. Since only
the vector part of the $(V+A)$ current contributes to the vector meson
production, $ \langle \eta |V-A|B\rangle \langle \omega|V+A | 0
\rangle =\langle \eta |V-A |B  \rangle \langle \omega|V-A|0 \rangle$,  that is
\beq
 F_{e}^{P1}=F_{e} \; .
\eeq
For other cases, one needs to do a Fierz transformation  for the
corresponding operators to get the right flavor and color structure for
factorization to work. We may get $(S+P)(S-P)$ operators from
$(V-A)(V+A)$ ones. Because neither scalar nor the pseudo-scalar
density give contribution to a vector meson production,
$\langle\omega |S+P |0 \rangle=0$, we get $F_{e}^{P2}= 0$.

For the non-factorizable diagrams 1(c) and 1(d), all
three meson wave functions are involved. The integration of $b_3$
can be performed using the $\delta$ function $\delta(b_3-b_1)$, leaving
only integration of $b_1$ and $b_2$.
The decay amplitude for $(V-A)(V-A)$ and $(V-A)(V+A)$
operators can be written as
\beq
M_{e}&=&- M_{e}^{P2}= \frac{16 }{\sqrt{3}}G_F\pi C_F m_B^4
\int_{0}^{1}d x_{1}d x_{2}\,d x_{3}\,\int_{0}^{\infty} b_1d b_1 b_2d
b_2\, \pb(x_1,b_1)\phi_\omega(x_2,b_2) \non
 & &\times
\left \{\left[2x_3\re\phi_\eta^T(x_3,b_1)-x_3
\phi_\eta^A(x_3,b_1)\right ]\right.\non
 & & \cdot \alpha_s(t_f) h_f(x_1,x_2,x_3,b_1,b_2)\exp[-S_{cd}(t_f)] \} , \\
M_e^{P1}&=& - \frac{32} {\sqrt{3}} G_F \pi C_F r_\omega
m_B^4 \int_{0}^{1}d x_{1}d x_{2}\,d x_{3}\,\int_{0}^{\infty} b_1d
b_1 b_2d b_2\, \phi_B(x_1,b_1) \non & &\cdot \left\{ \left[ x_2
\phi_\eta^A(x_3, b_1)
\left(\phi_\omega^s(x_2,b_2)-\phi_\omega^t(x_2,b_2)\right) + r_\eta
\left( (x_2+x_3)
\left(\phi_\eta^p(x_3,b_1)\right.\right.\right.\right. \non
 && \left.\left.\left.\left.\cdot \phi_\omega^s(x_2,b_2)
 + \phi_\eta^T(x_3,b_1)\phi_\omega^t(x_2,b_2)\right)
 +(x_3-x_2)\left(\phi_\eta^P(x_3,b_1)\phi_\omega^t(x_2,b_2)\right.\right.\right.\right.
 \non
 && \left.\left.\left.\left.+\phi_\eta^T(x_3,b_1)\phi_\omega^s(x_2,b_2)\right)\right)\right]
 \alpha_s(t_f)h_f(x_1,x_2,x_3,b_1,b_2)\exp[-S_{cd}(t_f)]\right \}
\;.
\eeq

For the non-factorizable annihilation diagrams 1(e)
and 1(f), there are three kinds of decay amplitudes.
For the  $(V-A)(V-A)$ operators, the decay amplitude $M_a$ is
 \beq
M_a&=& \frac{16} {\sqrt{3}} \pi G_F C_F  m_B^4 \int_{0}^{1}d x_{1}d x_{2}\,d
x_{3}\,\int_{0}^{\infty} b_1d b_1 b_2d b_2\, \phi_B(x_1,b_1)
 \non
 & &\cdot \left\{ \left[ r_\omega r_\eta (x_3-x_2)\left[\phi_\eta^P(x_3,b_2)\phi_\omega^t(x_2,b_2)+
\phi_\eta^T(x_3,b_2)\, \phi_\omega^s(x_2,b_2)\right ] \right.\right.
 \non
 &&
\left.\left. + r_\omega r_\eta (x_2+x_3)  \left [ \phi_\eta^P(x_3,b_2)\phi_\omega^s(x_2,b_2)+
\phi_\eta^T(x_3,b_2)\phi_\omega^t(x_2,b_2)\right ] \right.\right.
 \non
 &&\left. \left. + x_3 \phi_\omega(x_2,b_2)\phi_\eta^A(x_3,b_2) \right] \cdot \alpha_s(t_f^4)
h_f^4(x_1,x_2,x_3,b_1,b_2)\exp[-S_{ef}(t_f^4)]\right.
 \non
&&\left.-\left[
r_\omega  r_\eta (x_2-x_3)\left [\phi_\eta^P(x_3,b_2) \phi_\omega^t(x_2,b_2)
 +\phi_\eta^T(x_3,b_2) \phi_\omega^s(x_2,b_2)\right ]  \right.\right.
 \non
  && \left.\left.
 + r_\omega  r_\eta \left [ (2+x_2+x_3)\phi_\eta^P(x_3,b_2)\phi_\omega^s(x_2,b_2)
 -(2-x_2-x_3)\phi_\eta^T(x_3,b_2) \phi_\omega^t(x_2,b_2)\right]  \right.\right.
 \non
 &&\left.\left. +  x_2 \phi_\omega(x_2,b_2)\phi_\eta^A(x_3,b_2) \right]
\cdot  \alpha_s(t_f^3)
 h_f^3(x_1,x_2,x_3,b_1,b_2)\exp[-S_{ef}(t_f^3)]\right \}\; .
 \eeq

For the $(V-A)(V+A)$ and $(S-P)(S+P)$ operators, we have
\beq
 M_a^{P1} &=&\frac{16} {\sqrt{3}} G_F \pi C_F
m_B^4 \int_{0}^{1}d x_{1}d x_{2}\,d x_{3}\,\int_{0}^{\infty} b_1d
b_1 b_2d b_2\, \phi_B(x_1,b_1)
 \non
& &\cdot \left\{ \left[x_2 r_\omega \phi_\eta^A(x_3, b_2) \left
(\phi_\omega^s(x_2,b_2)+\phi_\omega^t(x_2,b_2) \right) - x_3 r_\eta
\left(\phi_\eta^P(x_3,b_2)+\phi_\eta^T(x_3,b_2) \right)
\right.\right.
 \non
 & &\left. \cdot \phi_\omega(x_2,b_2)] \cdot \alpha_s(t_f^4)h_f^4(x_1,x_2,x_3,b_1,b_2)
\exp[-S_{ef}(t_f^4)]\right.
 \non
& &\left.+\left[(2-x_2)r_\omega\phi_\eta^A(x_3, b_2) \left
(\phi_\omega^s(x_2,b_2)+\phi_\omega^t(x_2,b_2) \right)- (2-x_3)
r_\eta\left(\phi_\eta^P(x_3,b_2) \right.\right.\right.
 \non
 & &\left.\left.\left.
+\phi_\eta^T(x_3,b_2) \right)
 \phi_\omega(x_2,b_2) \right] \cdot \alpha_s(t_f^3)h_f^3(x_1,x_2,x_3,b_1,b_2)\exp[-S_{ef}(t_f^3)]\}\;
\right. , \\
 M_a^{P2}&=& -\frac{16}
{\sqrt{3}} \pi G_F C_F  m_B^4 \int_{0}^{1}d x_{1}d x_{2}\,d
x_{3}\,\int_{0}^{\infty} b_1d b_1 b_2d b_2\, \phi_B(x_1,b_1)
 \non
 & &\cdot \left\{ \left[x_2 \phi_\omega(x_2,b_2)\phi_\eta^A(x_3,b_2)+r_\omega
r_\eta
\left((x_2-x_3)\left(\phi_\eta^P(x_3,b_2)\phi_\omega^t(x_2,b_2)+
\phi_\eta^T(x_3,b_2)\right.\right.\right.\right.
 \non
 &&
 \left.\left.\left.\left.\cdot \phi_\omega^s(x_2,b_2)\right)+(x_2+x_3)
 \left(\phi_\eta^P(x_3,b_2)\phi_\omega^s(x_2,b_2)+
\phi_\eta^T(x_3,b_2)\phi_\omega^t(x_2,b_2)\right) \right)
\right]\right.
 \non
 &&\left. \cdot \alpha_s(t_f^4)
h_f^4(x_1,x_2,x_3,b_1,b_2)\exp[-S_{ef}(t_f^4)]\right.
 \non
&&\left. -\left[x_3 \phi_\omega(x_2,b_2)\phi_\eta^A(x_3,
b_2)+r_\omega
 r_\eta\left((x_3-x_2)\left(\phi_\eta^P(x_3,b_2) \phi_\omega^t(x_2,b_2)
 +\phi_\eta^T(x_3,b_2)\right.\right.\right.\right.
 \non
  && \left.\left.\left.\left.\cdot
 \phi_\omega^s(x_2,b_2)\right)+(2+x_2+x_3)\phi_\eta^P(x_3,b_2)\phi_\omega^s(x_2,b_2)
 -(2-x_2-x_3)\phi_\eta^T(x_3,b_2)\right.\right.\right.
 \non
 &&\left.\left.\left.\cdot \phi_\omega^t(x_2,b_2) \right)
\right]
\cdot  \alpha_s(t_f^3)
 h_f^3(x_1,x_2,x_3,b_1,b_2)\exp[-S_{ef}(t_f^3)]\right \}\; .
 \eeq

The factorizable annihilation diagrams 1(g) and 1(h)
involve only $\eta$ and $\omega$ wave functions. There are also
three kinds of decay amplitudes: $F_{a}$ is
for $(V-A)(V-A)$ type operators, $F_{a}^{P1}$ and $F_{a}^{P2}$  is for $(V-A)(V+A)$
and  $(S-P)(S+P)$ type operators, respectively:
\beq
 F_{a}&=&-F_{a}^{P1}\non
 &=&-4\sqrt{2}G_F\pi C_F m_B^4\int_{0}^{1}dx_{2}\,d x_{3}\,
\int_{0}^{\infty} b_2d b_2b_3d b_3 \, \left\{ \left[x_3
\pe^A(x_3,b_3) \phi_\omega(x_2,b_2)\right.\right.\non
&&\left.\left.+2r_\omega\re((x_3+1)\pep(x_3,b_3)+(x_3-1)
\pet(x_3,b_3)) \phi_\omega^s(x_2,b_2)\right] \right. \non && \left.
\quad \cdot \alpha_s(t_e^3) h_a(x_2,x_3,b_2,b_3)\exp[-S_{gh}(t_e^3)]
\right. \non && \left. -\left[ x_2
\pe^A(x_3,b_3)\phi_\omega(x_2,b_2) \right.\right.\non && \left.
\left. \quad +2 \re r_\omega ((x_2+1)\phi_\omega^s(x_2,b_2)+(x_2-1)
\phi_\omega^t(x_2,b_2) )\pep(x_3,b_3)\right] \right. \non &&\left.
\quad \cdot \alpha_s(t_e^4)
 h_a(x_3,x_2,b_3,b_2)\exp[-S_{gh}(t_e^4)]\right \}\; , \\
F_{a}^{P2}&=& -8\sqrt{2}G_F \pi C_F m_B^4 \int_{0}^{1}d
x_{2}\,d x_{3}\,\int_{0}^{\infty} b_2d b_2b_3d b_3 \,\non &&\times
\left\{ \left[x_3 \re (\pep(x_3, b_3)-\pet(x_3,
b_3))\phi_\omega(x_2,b_2)+2r_\omega \pe^A(x_3,b_3)
\phi_\omega^s(x_2,b_2) \right]\right.
 \non
&&\left.\quad \cdot \alpha_s(t_e^3)
h_a(x_2,x_3,b_2,b_3)\exp[-S_{gh}(t_e^3)]\right.
 \non
 &&\left.+\left[x_2r_\omega(\phi_\omega^s(x_2,b_2)-\phi_\omega^t(x_2,b_2))\pe^A(x_3,b_3)+2\re
\phi_\omega(x_2,b_2)\pep(x_3,b_3)\right] \right.\non &&\left.\quad
\cdot  \alpha_s(t_e^4)
 h_a(x_3,x_2,b_3,b_2)\exp[-S_{gh}(t_e^4)]\right\}\;.
 \eeq
In the above equations, we have assumed that $x_1 << (x_2, x_3)$. Since
the light quark momentum fraction $x_1$ in the $B$ meson is peaked at
the small region, while quark momentum fraction $x_3$ of $\eta$ is
peaked around $0.5$, this is not a bad approximation. The numerical
results also show that this approximation makes very little
difference in the final result.

\begin{figure}[tb]
\vspace{-3 cm}
\centerline{\epsfxsize=21 cm \epsffile{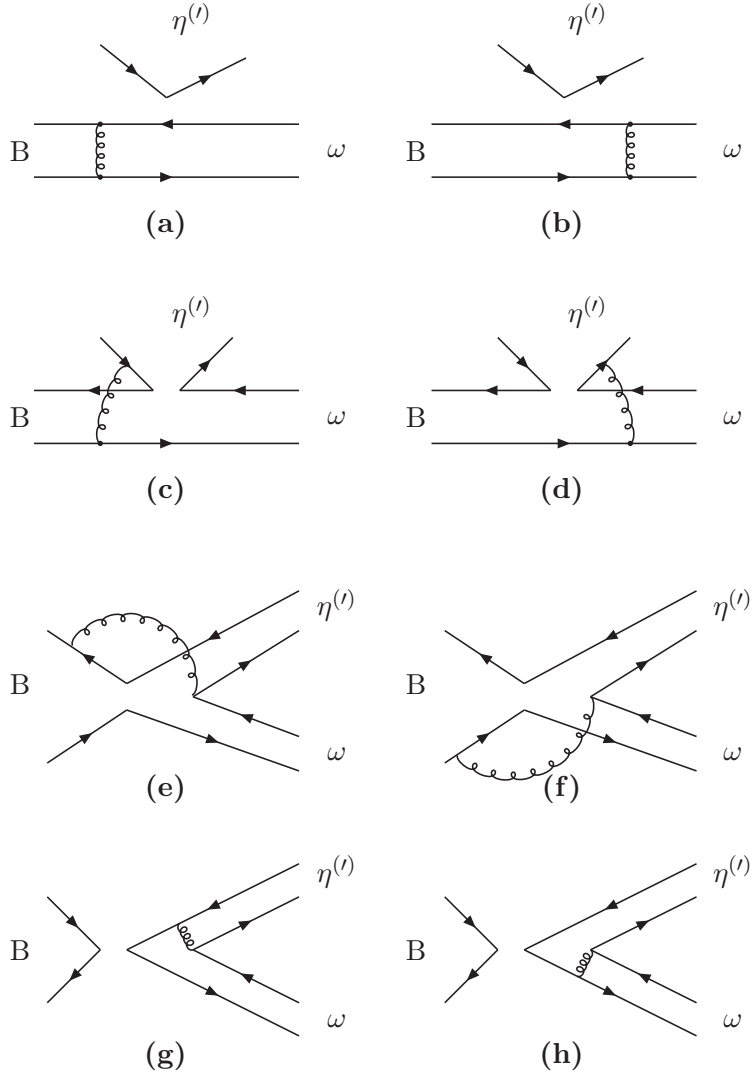}}
\vspace{-14cm}
\caption{ Feynman diagrams contributing to the $B\to \omega\eta^{(\prime)}$
 decays when the $\omega$ boson picks up the spectator quark.}
 \label{fig:fig2}
\end{figure}

By exchanging the position of $\etap$ and $\omega$ in Fig.~(\ref{fig:fig1}), one finds
the new Feynman diagrams are illustrated in Fig.~(\ref{fig:fig2}).
Just like the case of Fig.~(\ref{fig:fig1}), we firstly consider the factorizable diagrams
2(a) and 2(b). The decay amplitude $F_{e}$ induced by inserting the $(V-A)(V-A)$
operators is
\beq
 F_{e\omega}&=& 4 \sqrt{2} G_F \pi C_F m_B^4 \int_0^1 d x_{1} dx_{3}\,
\int_{0}^{\infty} b_1 db_1 b_3 db_3\, \phi_B(x_1,b_1)
 \non
& & \cdot \left\{ \left [(1+x_3) \phi_\omega (x_3, b_3)
+(1-2x_3)r_\omega (\phi_\omega^s (x_3, b_3)+\phi_\omega^t (x_3,b_3))
\right] \right.
 \non
 && \left. \cdot \alpha_s(t_e^1)
h_e(x_1,x_3,b_1,b_3)\exp[-S_{ab}(t_e^1)]
 \right.
  \non
&& \left. +2 r_\omega \phi_\omega^s (x_3, b_3) \cdot \alpha_s(t_e^2)
h_e(x_3,x_1,b_3,b_1)\exp[-S_{ab}(t_e^2)] \right\} \;, \label{eq:ab}
\eeq
The Fig.~2(a) and 2(b) are also the diagrams being used to extract
the form factor $A_0^{B\to \omega}$. According the definition in Ref. \cite{luyang},
 we can extract $A_0^{B\to \omega}$ from  Eq.(\ref{eq:ab}).

From  Fig.~ 2(a) and 2(b), it is easy to find the decay amplitude
$F_{e\omega}^{P1}$ and  $F_{e\omega}^{P2}$:
\beq
F_{e\omega}^{P1}&=& - F_{e\omega}, \\
F_{e\omega}^{P2}&=& 8 \sqrt{2} G_F \pi C_F r_\eta m_B^4 \int_{0}^{1}d x_{1}d
x_{3}\,\int_{0}^{\infty} b_1d b_1 b_3d b_3\, \phi_B(x_1,b_1) \non &
& \cdot
 \left\{ \left[ \phi_\omega (x_3, b_3)+r_\omega((x_3+2) \phi_\omega^s (x_3, b_3)
 - x_3 \phi_\omega^t (x_3, b_3))\right]\right.
\non
 & &\left. \cdot \alpha_s (t_e^1)  h_e
(x_1,x_3,b_1,b_3)\exp[-S_{ab}(t_e^1)]
  \right.  \non
& &\left.  + \left(x_1 \phi_\omega(x_3,b_3)+ 2 r_\omega
\phi_\omega^s (x_3, b_3)\right) \alpha_s (t_e^2)
 h_e(x_3,x_1,b_3,b_1)\exp[-S_{ab}(t_e^2)] \right\} \;.
\eeq

For the non-factorizable diagrams 2(c) and 2(d), the corresponding decay amplitudes are
\beq
 M_{e\omega}&=& - \frac{16} {\sqrt{3}} G_F \pi C_F m_B^4
\int_{0}^{1}d x_{1}d x_{2}\,d x_{3}\,\int_{0}^{\infty} b_1d b_1 b_2d
b_2\, \phi_B(x_1,b_1) \phi_\eta^A(x_2,b_2) \non
 & &\cdot
\left \{ x_3 \left[\phi_\omega(x_3,b_1)-2 r_\omega
\phi_\omega^t(x_3,b_1)\right]
 \alpha_s(t_f)
 h_f(x_1,x_2,x_3,b_1,b_2)\exp[-S_{cd}(t_f)]\right \},\ \
\eeq
and
\beq
  M_{e\omega}^{P1}= 0,\quad M_{e\omega}^{P2}=M_{e\omega}.
 \eeq

For the non-factorizable annihilation diagrams 2(e) and 2(f), again
all three wave functions are involved. the three kinds of
decay amplitudes are of the form
\beq
M_{a\omega}&=& \frac{16} {\sqrt{3}} G_F \pi C_F m_B^4 \int_{0}^{1}d
x_{1}d x_{2}\,d x_{3}\,\int_{0}^{\infty} b_1d b_1 b_2d b_2\,
\phi_B(x_1,b_1) \non && \cdot \left\{ \left[ x_3 \phi_\omega(x_3,
b_2) \phi_\eta^A(x_2,b_2) + r_\omega r_\eta \left( (x_3-x_2)
\left(\phi_\eta^P(x_2,b_2) \phi_\omega^t (x_3,b_2)+\phi_\eta^T
(x_2,b_2) \right.\right.\right.\right.
 \non
 &&\left. \left.\left.\left. \cdot \phi_\omega^s(x_3,b_2) \right)+(x_3+x_2)
 \left(\phi_\eta^P(x_2,b_2) \phi_\omega^s(x_3,b_2)
   +\phi_\eta^T (x_2,b_2) \phi_\omega^t(x_3,b_2)\right)\right)\right]\right.
   \non
&&\left.\cdot  \alpha_s(t_f^4)
h_f^4(x_1,x_2,x_3,b_1,b_2)\exp[-S_{ef}(t_f^4)]~-~\left[x_2
\phi_\omega(x_3,b_2) \phi_\eta^A(x_2,b_2)\right.\right. \non
 &&\left.\left.
 + r_\omega r_\eta \left( (x_2-x_3)\left(\phi_\eta^P(x_2,b_2)\phi_\omega^t(x_3,b_2)
 +\phi_\eta^T(x_2,b_2)\phi_\omega^s(x_3,b_2)\right)~+~r_\omega r_\eta
 \right.\right.\right.
\non && \left.\left.\left.\cdot
\left((2+x_2+x_3)\phi_\eta^P(x_2,b_2) \phi_\omega^s(x_3,b_2)
  - (2-x_2-x_3)\phi_\eta^T(x_2,b_2) \phi_\omega^t(x_3,b_2)\right)\right) \right]\right.
 \non
 & &\left. \cdot
 \alpha_s(t_f^3)h_f^3(x_1,x_2,x_3,b_1,b_2)\exp[-S_{ef}(t_f^3)] \right\}
 \;,
\eeq
 \beq
 M_{a\omega}^{P1}&=& -\frac{16} {\sqrt{3}} G_F \pi C_F m_B^4
 \int_{0}^{1}d x_{1}d x_{2}\,d x_{3}\,\int_{0}^{\infty} b_1 db_1 b_2 db_2\; \phi_B(x_1,b_1)
 \non
& &\cdot \left\{ \left[ x_2 r_\eta \phi_\omega(x_3, b_2)
\left (\phi_\eta^P(x_2,b_2)+\phi_\eta^T(x_2,b_2) \right) - x_3 r_\omega
\left(\phi_\omega^s(x_3,b_2)+\phi_\omega^t(x_3,b_2) \right)
\right.\right.
 \non
 & &\left. \left. \cdot \phi_\eta^A(x_2,b_2) \right ] \cdot \alpha_s(t_f^4)h_f^4(x_1,x_2,x_3,b_1,b_2)
\exp[-S_{ef}(t_f^4)] \right.
 \non
& &\left. + \left[(2-x_2)r_\eta \phi_\omega(x_3, b_2)
\left (\phi_\eta^P(x_2,b_2)+\phi_\eta^T(x_2,b_2) \right)- (2-x_3)
r_\omega\left( \phi_\omega^s(x_3,b_2) \right.\right.\right.
 \non
 & &\left.\left.\left. +\phi_\omega^t(x_3,b_2) \right)
 \phi_\eta^A(x_2,b_2) \right] \cdot \alpha_s(t_f^3)h_f^3(x_1,x_2,x_3,b_1,b_2)\exp[-S_{ef}(t_f^3)]\right \},
\eeq
\beq
M_{a\omega}^{P2}&=& -\frac{16} {\sqrt{3}} \pi G_F C_F  m_B^4
\int_{0}^{1}d x_{1}d x_{2}\,d x_{3}\,\int_{0}^{\infty}
b_1d b_1 b_2d b_2\, \phi_B(x_1,b_1)
 \non
 & &\cdot \left\{ \left[x_2 \phi_\omega(x_3,b_2)\phi_\eta^A(x_2,b_2)+r_\omega
r_\eta
\left((x_2-x_3)\left(\phi_\eta^P(x_2,b_2)\phi_\omega^t(x_3,b_2)+
\phi_\eta^T(x_2,b_2)\right.\right.\right.\right.
 \non
 &&
 \left.\left.\left.\left.\cdot \phi_\omega^s(x_3,b_2)\right)+(x_2+x_3)
 \left(\phi_\eta^P(x_2,b_2)\phi_\omega^s(x_3,b_2)+
\phi_\eta^T(x_2,b_2)\phi_\omega^t(x_3,b_2)\right) \right)
\right]\right.
 \non
 &&\left. \cdot \alpha_s(t_f^4)
h_f^4(x_1,x_2,x_3,b_1,b_2)\exp[-S_{ef}(t_f^4)]\right.
 \non
&&\left. -\left[x_3 \phi_\omega(x_3,b_2)\phi_\eta^A(x_2,
b_2)+r_\omega
 r_\eta\left((x_3-x_2)\left(\phi_\eta^P(x_2,b_2) \phi_\omega^t(x_3,b_2)
 +\phi_\eta^T(x_2,b_2)\right.\right.\right.\right.
 \non
  && \left.\left.\left.\left.\cdot
 \phi_\omega^s(x_3,b_2)\right)+(2+x_2+x_3)\phi_\eta^P(x_2,b_2)\phi_\omega^s(x_3,b_2)
 -(2-x_2-x_3)\phi_\eta^T(x_2,b_2)\right.\right.\right.
 \non
 &&\left.\left.\left.\cdot \phi_\omega^t(x_3,b_2) \right)
\right]
\cdot  \alpha_s(t_f^3)  h_f^3(x_1,x_2,x_3,b_1,b_2)\exp[-S_{ef}(t_f^3)]\right \}.
\eeq

For the factorizable annihilation diagrams 2(g) and 2(h), we have
\beq
F_{a\omega}&=& -F_{a},\quad F_{a\omega}^{P1}= -F_{a}^{P1},\quad F_{a\omega}^{P2}=
 -F_{a}^{P2}.
\eeq

Following the same procedure, it is straightforward to calculate the decay amplitudes for the
$B \to \phi \etap$ decays. It is worth of mentioning that all the eight Feynman diagrams in Fig.~1
after the replacements of the $\omega$ meson by the $\phi$ meson will contribute to $B \to \phi \etap$
decays; but Figs.2(a)-2(d) could not contribute to the same decay since $\phi=-s\bar{s}$ in the ideal
mixing scheme of $\omega-\phi$ system.


Combining the contributions from the different diagrams, the total decay
amplitude for $B^0 \to \omega\eta $ can be written as
\beq
\sqrt{2}{\cal M}(\omega\eta) &=& F_{e}F_1(\phi) f_{\omega}\left[ \xi_u \left( C_1 +
\frac{1}{3}C_2\right)\right.
 \non
& &\left.-\xi_t
\left(\frac{7}{3}C_3+\frac{5}{3}C_4+2C_{5}+\frac{2}{3}C_{6}
+\frac{1}{2}C_7+\frac{1}{6}C_8+\frac{1}{3}C_9
 -\frac{1}{3} C_{10}\right)\right ]
 \non
& & + M_{e}F_1(\phi)\left [ \xi_uC_2-\xi_t
\left(C_3+2C_4-2C_6-\frac{1}{2}C_8-\frac{1}{2}C_9
+\frac{1}{2}C_{10}\right)\right ]
  \non && +F_{e\omega}
 \left[ \xi_u \left( C_1 +
\frac{1}{3}C_2\right)f_{\eta}^{d}\right.
 \non
& &\left.-\xi_t
\left(\frac{7}{3}C_3+\frac{5}{3}C_4-2C_{5}-\frac{2}{3}C_{6}
-\frac{1}{2}C_7-\frac{1}{6}C_8+\frac{1}{3}C_9
 -\frac{1}{3} C_{10}\right)f_{\eta}^{d}\right.\non & &\left.
 -\xi_t\left(C_3+\frac{1}{3}C_4-C_5-\frac{1}{3}C_6
 +\frac{1}{2}C_7+\frac{1}{6}C_8-\frac{1}{2}C_9-\frac{1}{6}C_{10}\right)f_{\eta}^{s}\right]
 \non
& & +F_{e\omega}^{P_2}\left[-\xi_t\left(\frac{1}{3}C_5+C_6-
\frac{1}{6}C_7-\frac{1}{2}C_8\right)f_{\eta}^{d}\right] \non &
&+M_{e\omega}F_1(\phi)\left [ \xi_uC_2-\xi_t
\left(C_3+2C_4+2C_6+\frac{1}{2}C_8-\frac{1}{2}C_9
+\frac{1}{2}C_{10}\right)\right ]
  \non && + M_{e\omega}F_2(\phi)\left [-\xi_t
\left(C_4+C_6-\frac{1}{2}C_8-\frac{1}{2}C_{10}\right)\right ]
  \non &&+
\left(M_{a}+M_{a\omega}\right) F_1(\phi)\left[ \xi_uC_{2}-\xi_t
\left(C_3+2C_4-\frac{1}{2}C_9 +\frac{1}{2}C_{10}\right)\right ] \non
&&
 -\left(M_{e}^{P_1}\,+M_{a}^{P_1}\,+M_{a\omega}^{P_1}\,\right)F_1(\phi)
\xi_t\,\left(C_{5}-\frac{1}{2}C_{7}\right)
 \non && -
 \left(M_{a}^{P_2}\,+
 M_{a\omega}^{P_2}\,\right) F_1(\phi)\xi_t\,
\left(2C_6+\frac{1}{2}C_8\right) , \label{eq:m1}
\eeq
where $\xi_u = V_{ub}^*V_{ud}$, $\xi_t = V_{tb}^*V_{td}$, and
\beq
F_1(\phi)=\frac{\cos\phi}{\sqrt{2}},
\quad F_2(\phi)=-\sin\phi ,
\label{eq:f1f2}
\eeq
are the mixing factors. The Wilson coefficients
$C_i=C_i(t)$ in Eq.~(\ref{eq:m1}) should be calculated at the
appropriate scale $t$ using equations as given in the Appendices of
Ref.~\cite{luy01}.

By doing the replacements of $f_\omega \to f_{\phi}$ and   $\phi^{A,P,T}_{\omega} \to
\phi^{A,P,T}_{\phi}$, one obtains  the decay amplitude for $B^0 \to \phi \eta$ decay:
\beq
{\cal M}(\phi\eta) &=& F_{e}\;  \xi_t \left(C_3+\frac{1}{3}C_4+C_{5}+\frac{1}{3}C_{6}
-\frac{1}{2}C_7-\frac{1}{6}C_8-\frac{1}{2}C_9
 -\frac{1}{6} C_{10}\right) F_1(\phi) \non
& & + M_{e} \; \xi_t \; \left(C_4-C_6+\frac{1}{2}C_8
-\frac{1}{2}C_{10}\right) \; F_1(\phi)
  +\left( M_{a}+ M_{a\phi}\right) \; \xi_t
\left(C_4-\frac{1}{2}C_{10}\right) F_2(\phi) \non &&
+\left(M_{a}^{P_2}+M_{a\phi}^{P_2}\right) \;  \xi_t\,
\left(C_6-\frac{1}{2}C_8\right) F_2(\phi). \label{eq:m2}
\eeq

The complete decay amplitude  ${\cal M}(\omega \etapr)$  and ${\cal
M}(\phi \etapr)$ can be obtained easily from Eq.(\ref{eq:m1}) and
Eq.(\ref{eq:m2}) by the following replacements
\beq
f_\eta^{d},\; f_\eta^s &\longrightarrow & f_{\eta^\prime}^d, \; f_{\eta^\prime}^s,
\non F_1(\phi) &\longrightarrow & F'_1(\phi) =
\frac{\sin\phi}{\sqrt{2}}, \non
F_2(\phi) &\longrightarrow & F'_2(\phi)
=\cos\phi.
\eeq

\section{Numerical results and Discussions}\label{sec:n-d}

In this section, we will present the numerical results of the branching ratios and CP violating asymmetries
for those considered decay modes.

\subsection{Input parameters} \label{sec:input}

We show here the input parameters  to be used in the numerical calculations.

The masses, decay constants, QCD scale  and $B^0$ meson lifetime are
\beq
 \Lambda_{\overline{\mathrm{MS}}}^{(f=4)} &=& 250 {\rm MeV}, \quad
 f_\pi = 130 {\rm MeV}, \quad f_B = 190 {\rm MeV}, \non
f_\phi &=& 237 {\rm MeV},\quad f_\phi^T = 220 {\rm MeV}, \quad M_W = 80.41{\rm  GeV},  \non
f_\omega&=& 195{\rm MeV},  \quad f_\omega^T =140{\rm MeV}, \quad M_\phi = 1.02 {\rm GeV},
\non M_\omega&=&0.78 {\rm GeV}, \quad m_\eta=547.5 {\rm  MeV}, \quad m_{\eta^\prime}=957.8 {\rm  MeV}, \non
 M_B &=& 5.2792 {\rm GeV}, \quad \tau_{B^{0}}=1.54\times10^{-12}{\rm s}.
 \label{para}
\eeq

For the Cabibbo-Kobayashi-Maskawa (CKM) matrix elements, here we adopt the Wolfenstein
parametrization for the CKM matrix up to $\mathcal{O}(\lambda^5)$ with the parameters
$\lambda=0.2272, A=0.818, \bar{\rho}=0.221$ and $\bar{\eta}=0.340$\cite{pdg06}.

From Eq.(\ref{eq:ab}) we find the numerical values of the corresponding form factors at zero
momentum transfer:
$A_0^{B \to \omega}(q^2=0)=0.35 \pm0.05(\omega_b) \label{eq:aff0}$
which are consistent with those given in \cite{sum1}.

\subsection{Branching ratios}

For $B^0 \to \omega \etap$ and $\phi \etap $ decays, the decay
amplitudes as given  in Eqs.~(\ref{eq:m1}) and (\ref{eq:m2}) can be
rewritten as
 \beq
{\cal M} &=& V_{ub}^*V_{ud} T -V_{tb}^* V_{td} P= V_{ub}^*V_{ud} T
\left [ 1 + z e^{ i ( \alpha + \delta ) } \right], \label{eq:ma}
\eeq
where
\beq
z=\left|\frac{V_{tb}^* V_{td}}{ V_{ub}^*V_{ud} }\right| \left|\frac{P}{T}\right|,
\qquad
\alpha = \arg \left[-\frac{V_{td}V_{tb}^*}{V_{ud}V_{ub}^*}\right]  \label{eq:zz}
\eeq
are the ratio of penguin to tree contributions and the weak
phase (one of the three CKM angles) respectively, and $\delta$ is the relative
strong phase between tree (T) and penguin (P) diagrams.
The CP-averaged branching ratio for $B^0 \to \omega
(\phi)\etap$ decays can be then written as
\beq
Br= (|{\cal M}|^2 +|\overline{\cal M}|^2)/2 =  \left| V_{ub}V_{ud}^* T \right| ^2 \left[1 +2 z\cos
\alpha \cos \delta +z^2 \right], \label{br}
\eeq
where the ratio $z$ and the strong phase $\delta$ have been defined in Eqs.(\ref{eq:ma})
and (\ref{eq:zz}).

Using  the wave functions and the input parameters as specified in
previous sections, it is straightforward to calculate the branching
ratios for the four considered decays. We find numerically that
\beq
Br(\ B^0 \to\omega\eta) &=& \left [2.7^{+0.8}_{-0.6}(\omega_b)
  \pm 0.7 (\alpha )\pm 0.3(a_{2})\right ] \times 10^{-7},
 \label{eq:brew}\\
Br(\ B^0 \to \omega\eta^{\prime}) &=& \left [0.75^{+0.19}_{-0.14}(
\omega_b)^{+0.25} _{-0.24}(\alpha ) ^{+0.19} _{-0.17}(a_{2}) \right
] \times 10^{-7}, \label{eq:brpw}
 \eeq
 The main errors are induced by the uncertainties of $\omega_b=0.4 \pm 0.04$ GeV,
 $\alpha =100^\circ \pm 20^\circ$ and $a_{2}=0.115 \pm 0.115$, respectively.

For the $B^0 \to \phi \etap$ decays, the decay amplitudes involve only the penguin (P)
diagrams, the branching ratios are:
\beq
 Br(\ B^0 \to\phi\eta) &=& \left [6.3^{+1.2}_{-1.0}(\omega_b)
 ^{+2.7}_{-1.0}(m_s)^{+1.4} _{-1.3}(a_{2})\right ] \times 10^{-9},
 \label{eq:brep}\\
Br(\ B^0 \to \phi\eta^{\prime}) &=& \left [7.3^{+1.5}_{-1.3}( \omega_b)^{+2.3} _{-0.9}(m_s)
^{+2.2} _{-2.0}(a_{2})\right ] \times 10^{-9}, \label{eq:brpp}
\eeq
The main errors are induced by the uncertainties of $\omega_b=0.4 \pm 0.04$
GeV, $m_s = 130 \pm 30$ MeV and $a_{2}=0.115 \pm 0.115$, respectively.

For the CP-averaged branching ratios of the considered four decays,
the PQCD predictions agree within errors with both the experimental upper limits as shown in
Eqs.(\ref{eq:exp1}-\ref{eq:ulimits}), and the theoretical predictions in the QCDF approach, for example,
as given in Ref.~\cite{bn03b}:
 \beq
 Br(\ B^0 \to\omega \eta) &=&\left ( 0.31 ^{+0.46}_{-0.27}\right ) \times 10^{-6},
\label{eq:br11}\\
 Br(\ B^0 \to \omega\eta^{\prime}) &=& \left ( 0.20 ^{+0.34}_{-0.18}\right ) \times
 10^{-6},\\
 Br(\ B^0 \to\phi \eta^{(\prime)}) &\approx& 1  \times 10^{-9},
 \label{eq:br24}
\eeq
where the individual errors as given in Refs.~\cite{bn03b} have
been added in quadrature. It is easy to see that the central values of the
pQCD predictions for $Br(B \to\phi\etap)$ are larger than the corresponding QCDF predictions,
although they are still consistent if the large theoretical uncertainties are taken into account.
The branching ratios  $Br(B \to \omega \etap)$ at the $10^{-7}$ level can be measured in the forthcoming
LHC experiments \cite{lhcb}. But it is very difficult to measure $B \to \phi \etap$ decays because of
their tiny ($\sim 10^{-9}$) branching ratio, if the pQCD predictions are true.

\begin{figure}[tb]
\centerline{\mbox{\epsfxsize=9cm\epsffile{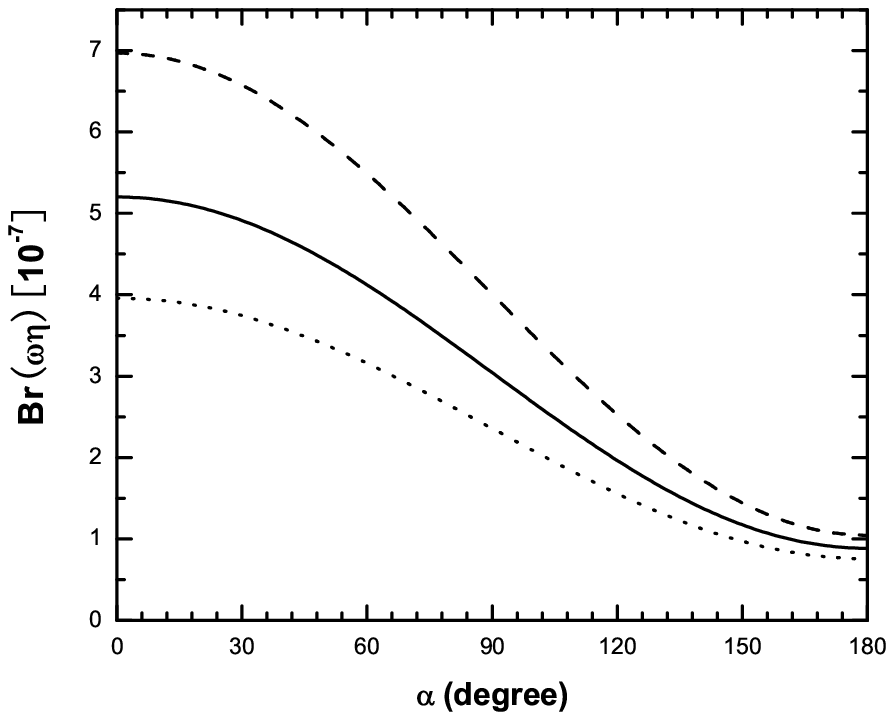}\epsfxsize=9cm\epsffile{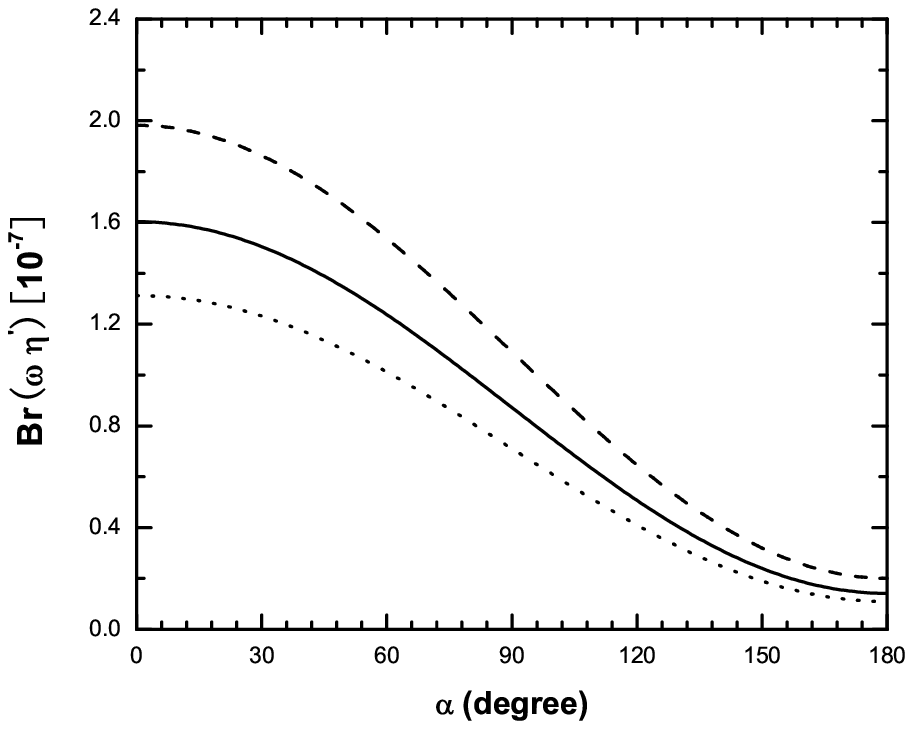}}}
\vspace{0.3cm}
\caption{The $\alpha$ dependence of the branching ratios (in unit of $10^{-7}$)
of $B^0\to \omega\eta$ and  $\omega\eta^{\prime}$ decays for
$\phi=39.3^\circ$,  $\omega_b=0.36 $ GeV (dashed curve), $0.40$ GeV (solid  curve)
and $0.44$ GeV (dotted curve).}  \label{fig:fig3}
\end{figure}

It is worth stressing  that the theoretical predictions in the pQCD
approach have large theoretical errors induced by
the still large uncertainties of many input parameters. In our
analysis, we considered the constraints on these parameters from
analysis of other well measured decay channels.
From numerical calculations, we get to know that the main errors come from the
uncertainty of $\omega_b$, $m_s$, $\alpha$ and $a_{2}^{\eta_{q(s)}}$.
Additionally, the final-state interactions remains unsettled in pQCD, which is non-perturbative
in nature but not universal.

In Fig.~\ref{fig:fig3} we show the parameter dependence of the pQCD
predictions for the branching ratios of $B \to \omega\eta$ and
$\omega \eta^\prime$ decays for  $\omega_b=0.4\pm 0.04$ GeV,  $\alpha=[0^\circ,180^\circ]$.
From the numerical results and the figures we observe that the pQCD
predictions are sensitive to the variations of $\omega_b$ and $\alpha$.

\subsection{CP-violating asymmetries }

Now we turn to the evaluations of the CP-violating asymmetries of $B
\to \omega\etap$ decays in pQCD approach. Because these decays are
neutral B meson decays, so we should consider the effects of
$B^0-\bar{B}^0$ mixing. For $B^0$ meson decays into a CP eigenstate
$f$, the time-dependent CP-violating asymmetry can be defined as
\beq
\frac{Br \left (\overline{B}^0(t) \to f \right) -
Br\left(B^0(t) \to f\right )}{ Br\left (\overline{B}^0(t) \to
f\right ) + Br\left (B^0(t) \to f\right ) } \equiv \acp^{dir} \cos
(\Delta m  \; t) + \acp^{mix} \sin (\Delta m \; t), \label{eq:acp-def}
\eeq
where $\Delta m$ is the mass difference
between the two $B_d^0$ mass eigenstates, $ t =t_{CP}-t_{tag} $ is
the time difference between the tagged $B^0$ ($\overline{B}^0$) and
the accompanying $\overline{B}^0$ ($B^0$) with opposite b flavor
decaying to the final CP-eigenstate $f_{CP}$ at the time $t_{CP}$.
The direct and mixing induced CP-violating asymmetries $\acp^{dir}$
and $\acp^{mix}$ can be written as
\beq
\acp^{dir}=\frac{ \left |\lambda_{CP}\right |^2 -1 } {1+|\lambda_{CP}|^2}, \qquad
A_{CP}^{mix}=\frac{ 2Im (\lambda_{CP})}{1+|\lambda_{CP}|^2},
\label{eq:acp-dm}
\eeq
where the CP-violating parameter $\lambda_{CP}$ is
\beq
\lambda_{CP} = \frac{ V_{tb}^*V_{td} \langle
f |H_{eff}| \overline{B}^0\rangle} { V_{tb}V_{td}^* \langle f
|H_{eff}| B^0\rangle} = e^{2i\alpha}\frac{ 1+z e^{i(\delta-\alpha)}
}{ 1+ze^{i(\delta+\alpha)} }. \label{eq:lambda2}
\eeq
Here the ratio $z$ and the strong phase $\delta$ have been defined previously. In
pQCD approach, since both $z$ and $\delta$ are calculable, it is
easy to find the numerical values of $\acp^{dir}$ and $\acp^{mix}$
for the considered decay processes.

By using the central values of the input parameters, one found the
pQCD predictions (in unit of $10^{-2}$) for the direct and mixing
induced CP-violating asymmetries of the considered decays
\beq
\acp^{dir}(B^0 \to \omega \eta)=-69.1^{+15.1}_{-13.4}(\alpha),\qquad
\acp^{mix}(B^0 \to \omega \eta)= +66.9^{+5.3}_{-15.8}(\alpha),\label{eq:acp1} \\
\acp^{dir}(B^0
\to\omega\eta^\prime)=+13.9^{+4.1}_{-3.5}(\alpha),\qquad
\acp^{mix}(B^0 \to\omega \eta^\prime)=+65.8^{+29.1}_{-55.2} (\alpha), \label{eq:acp2}
\eeq
where the dominant errors come from the variations of $\alpha=100^\circ \pm 20^\circ$.

\begin{figure}[tb]
\centerline{\mbox{\epsfxsize=9cm\epsffile{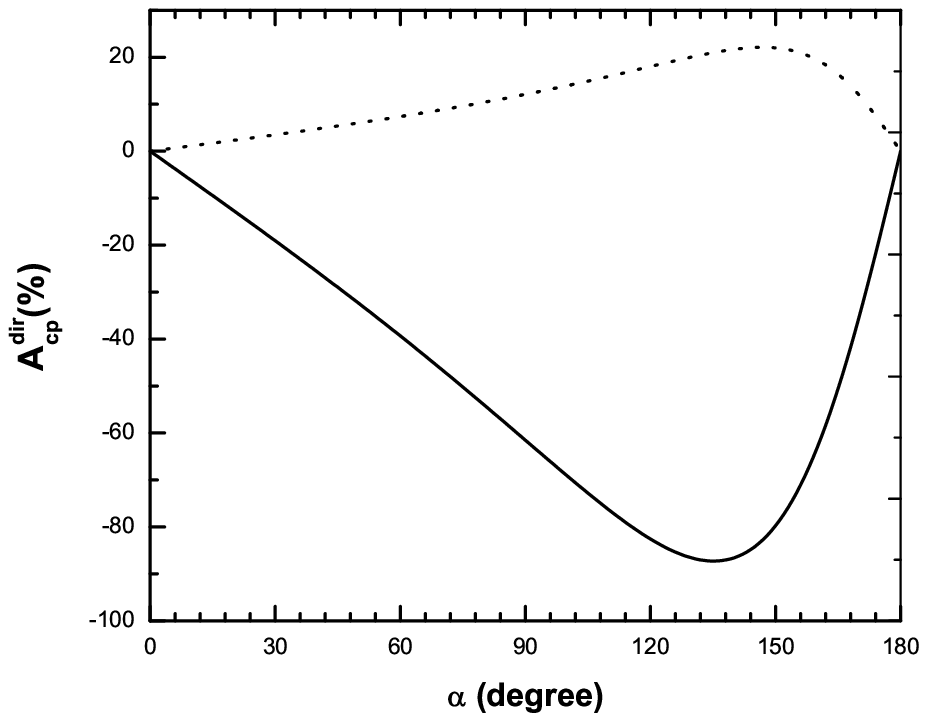}\epsfxsize=9cm\epsffile{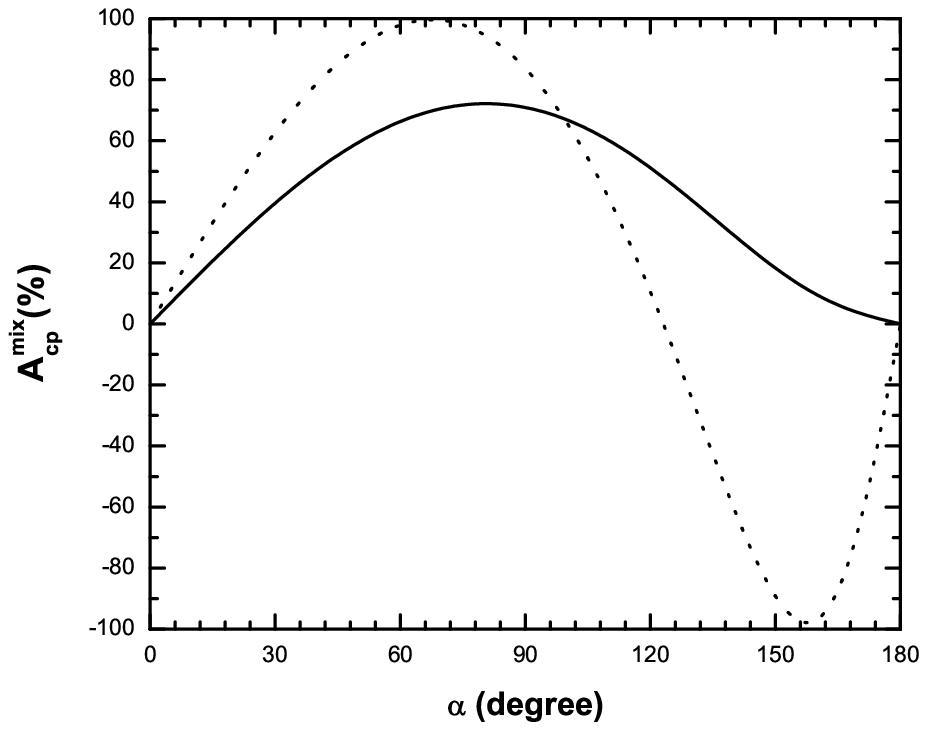}}}
\vspace{0.3cm} \caption{The direct and mixing-induced CP asymmetry
(in percentage) of $B^0\to \omega \eta$(solid curve)  and $\omega
\eta^{\prime}$(dotted curve) decay as a function of CKM angle $\alpha$.}
\label{fig:fig4}
\end{figure}

In Fig.~\ref{fig:fig4}, we show the $\alpha-$dependence of the pQCD
predictions for the direct and the mixing-induced CP-violating
asymmetry for $B^0 \to \omega\eta$(solid curve) and  $B^0 \to \omega
\eta^\prime$ (dotted curve) decay, respectively. The pQCD predictions in Eqs.~(\ref{eq:acp1},\ref{eq:acp2})
are consistent with the QCDF predictions due to very large uncertainties in the QCDF approach \cite{bn03b}.

If we integrate the time variable $t$, we will get the total CP
asymmetry for $B^0 \to \omega \etap$ decays (in units of $10^{-2}$):
\beq
\acp^{tot}(B^0 \to \omega\eta)&=&-11^{+12.1}_{-15.1}(\alpha), \\
\acp^{tot}(B^0 \to \omega\eta^\prime)&=& +40.6^{+11.8}_{-24.2}(\alpha) .
\eeq
For the $B^0 \to\phi \etap$ decay modes, however, there is no CP asymmetry, the reason is that
they involve only penguin contributions and one type of CKM elements.

It is worth of mentioning that although the CP-violating asymmetries of $B \to \omega \etap$ decays are
rather large in size, but it may be difficult to measure them because of the predicted small
branching ratios at $10^{-7}$ level \cite{lhcb}.

\subsection{Effects of possible gluonic component of $\eta^\prime$}

Up to now, we have not considered the possible contributions to the
branching ratios and CP-violating asymmetries of $B \to \omega(\phi)
\etap$ decays induced by the possible gluonic component of
$\eta$ and $\eta^\prime$ meson. For $\eta$ meson, one generally believe that its gluonic component
is negligibly small. For $\eta^\prime$ meson, however, a large gluonic component is generally expected
because of the observed very large $B \to K \eta^\prime$ branching ratio.

When a non-zero gluonic component exist in $\etapr$ meson, an additional
decay amplitude ${\cal M}'$ will be produced. Such decay amplitude may construct or
destruct with the ones from the $q\bar{q}$ ($q=u,d,s$) components of
$\eta^\prime$, the branching ratios of the decays in question may be increased or decreased accordingly.

In Ref.~\cite{li0609}, by employing the pQCD factorization approach,
 Charng, Kurimoto and Li calculated the flavor-singlet contribution to
 the $B \to \etap$ transition form factors
induced by the Feynman diagrams with the two gluons emitted from the light
quark of the B meson (see Fig.~1 of Ref.~\cite{li0609}), the dominant gluonic
contribution is negligible for $B \to \eta$ form factor, and also small (less than $5\%$)
for the $B \to \eta^\prime$ ones \cite{li0612}.

Following the procedure of Ref.~\cite{li0609} and using the formulae given there,
we also calculate the gluonic contributions to $B \to \omega (\phi) \eta^\prime$ decays.
If we add the gluonic contribution to the form factor $F_0^{B\to\eta^\prime}$ but keeping
all other inputs in their default central values,  we find numerically that
\beq
Br(\ B^0\to \omega\eta^{\prime}) &=& 0.82\times 10^{-7}, \\
Br(\ B^0 \to \phi\eta^{\prime}) &=& 6.5 \times 10^{-9}.
\eeq
One can see that the variation of the central values of the pQCD predictions induced by the inclusion of
gluonic contribution is only about $\pm 10\%$, much smaller than the theoretical uncertainties from
the variations of input parameters $\omega_b$, $m_s$, or $\alpha$.

\section{summary }

In this paper, we calculated the branching ratios of $B^0 \to
\omega\etap $, $ \phi\etap$ decays and CP-violating asymmetries of
$B^0 \to \omega\etap $ decays at the leading order by using the pQCD factorization approach.

Besides the usual factorizable diagrams, the non-factorizable and
annihilation diagrams are also calculated analytically. Although
the non-factorizable and annihilation contributions are
sub-leading for the branching ratios of the considered decays, but
they are not negligible. Furthermore these diagrams provide the
necessary strong phase required by a non-zero CP-violating  asymmetry for the considered decays.

After calculating all the diagrams, we found the branching ratios of
the four related decays are very small, $Br(B \to \omega\etap)$ are
at the order of $O(10^{-7})$ and $Br(B \to \phi\etap)$ are around $10^{-9}$.
We also predict the direct and mixing-induced CP asymmetries of
the $B \to \omega\etap$. Moreover, we compare our results with the
current experimental data and the theoretical predictions in QCDF approach.
In short, we found that
\begin{itemize}
\item
For the CP-averaged branching ratios of the considered decay modes, the pQCD predictions are
\beq
Br(\ B^0 \to\omega\eta) &=& \left ( 2.7^{+1.1}_{-1.0} \right ) \times 10^{-7},  \label{eq:brew2}\\
Br(\ B^0 \to \omega\eta^{\prime}) &=& \left ( 0.75^{+0.37}_{-0.33}  \right ) \times 10^{-7},
\label{eq:brpw2}\\
Br(\ B^0 \to\phi\eta) &=& \left ( 6.3^{+3.3}_{-1.9} \right ) \times 10^{-9}, \label{eq:brep2}\\
Br(\ B^0 \to \phi\eta^{\prime}) &=& \left ( 7.3^{+3.5}_{-2.6} \right ) \times 10^{-9}, \label{eq:brpp2}
\eeq
where the various errors as specified in Eqs.~(\ref{eq:brew})-(\ref{eq:brpp})
have been added in quadrature. The inclusion of the gluonic contribution can change
the branching ratios of $B \to \omega\eta^\prime, \phi \eta^\prime )$ decays by
about $10\%$ in magnitude.

\item
The pQCD predictions for the CP-violating asymmetries of $B \to \omega \etap$ decays are large in size.

\item
The major theoretical errors are induced by the uncertainties of the input parameters
$\omega_{B}$, the mass $m_s$, the CKM angle $\alpha$ and the Gegenbauer moment $a_2^{\eta_{q(s)}}$.

\end{itemize}

\begin{acknowledgments}

We are very grateful to Cai-Dian L\"u, Ying Li, Xin Liu and Huisheng Wang for helpful discussions.
This work is partly supported  by the National Natural Science Foundation of China under Grant
No.10575052, and by the Specialized Research Fund for the doctoral Program of higher
education (SRFDP) under Grant No.~20050319008.

\end{acknowledgments}


\begin{appendix}

\section{Related Functions }\label{sec:app1}

We show here the function $h_i$ appearing in the expressions of the decay amplitudes in
section \ref{sec:p-c}, coming from the Fourier transformations  of the function $H^{(0)}$,
\beq
 h_e(x_1,x_3,b_1,b_3)&=&
 K_{0}\left(\sqrt{x_1 x_3} m_B b_1\right)
 \left[\theta(b_1-b_3)K_0\left(\sqrt{x_3} m_B
b_1\right)I_0\left(\sqrt{x_3} m_B b_3\right)\right.
 \non
& &\;\left. +\theta(b_3-b_1)K_0\left(\sqrt{x_3}  m_B b_3\right)
I_0\left(\sqrt{x_3}  m_B b_1\right)\right] S_t(x_3), \label{he1}
\eeq
 \beq
 h_a(x_2,x_3,b_2,b_3)&=&
 K_{0}\left(i \sqrt{x_2 x_3} m_B b_2\right)
 \left[\theta(b_3-b_2)K_0\left(i \sqrt{x_3} m_B
b_3\right)I_0\left(i \sqrt{x_3} m_B b_2\right)\right.
 \non
& &\;\;\;\;\left. +\theta(b_2-b_3)K_0\left(i \sqrt{x_3}  m_B
b_2\right) I_0\left(i \sqrt{x_3}  m_B b_3\right)\right] S_t(x_3),
\label{he3} \eeq
 \beq
 h_{f}(x_1,x_2,x_3,b_1,b_2) &=&
 \biggl\{\theta(b_2-b_1) \mathrm{I}_0(M_B\sqrt{x_1 x_3} b_1)
 \mathrm{K}_0(M_B\sqrt{x_1 x_3} b_2)
 \non
&+ & (b_1 \leftrightarrow b_2) \biggr\}  \cdot\left(
\begin{matrix}
 \mathrm{K}_0(M_B F_{(1)} b_2), & \text{for}\quad F^2_{(1)}>0 \\
 \frac{\pi i}{2} \mathrm{H}_0^{(1)}(M_B\sqrt{|F^2_{(1)}|}\ b_2), &
 \text{for}\quad F^2_{(1)}<0
\end{matrix}\right),
\label{eq:pp1}
 \eeq
\beq
h_f^3(x_1,x_2,x_3,b_1,b_2) &=& \biggl\{\theta(b_1-b_2)
\mathrm{K}_0(i \sqrt{x_2 x_3} b_1 M_B)
 \mathrm{I}_0(i \sqrt{x_2 x_3} b_2 M_B)+(b_1 \leftrightarrow b_2) \biggr\}
 \non
& & \cdot
 \mathrm{K}_0(\sqrt{x_1+x_2+x_3-x_1 x_3-x_2 x_3}\ b_1 M_B),
 \label{eq:pp4}
\eeq
 \beq
 h_f^4(x_1,x_2,x_3,b_1,b_2) &=&
 \biggl\{\theta(b_1-b_2) \mathrm{K}_0(i \sqrt{x_2 x_3} b_1 M_B)
 \mathrm{I}_0(i \sqrt{x_2 x_3} b_2 M_B)
 \non
&+& (b_1 \leftrightarrow b_2) \biggr\} \cdot \left(
\begin{matrix}
 \mathrm{K}_0(M_B F_{(2)} b_1), & \text{for}\quad F^2_{(2)}>0 \\
 \frac{\pi i}{2} \mathrm{H}_0^{(1)}(M_B\sqrt{|F^2_{(2)}|}\ b_1), &
 \text{for}\quad F^2_{(2)}<0
\end{matrix}\right), \label{eq:pp3}
\eeq
where $J_0$ is the Bessel function, $K_0$ and $I_0$ are
the modified Bessel functions $K_0 (-i x) = -(\pi/2) Y_0 (x) + i
(\pi/2) J_0 (x)$, $H_0^{(1)}(z)=J_0(z) + i Y_0(z)$, and the $F_{(j)}$ are defined by
\beq
F^2_{(1)}&=&(x_1 -x_2) x_3\;,\\
F^2_{(2)}&=&(x_1-x_2) x_3\;\;.
 \eeq

The threshold resummation form factor $S_t(x_i)$ is adopted from
Ref.~\cite{kurimoto}
\beq
S_t(x)=\frac{2^{1+2c} \Gamma
(3/2+c)}{\sqrt{\pi} \Gamma(1+c)}[x(1-x)]^c,
\eeq
where the parameter $c=0.3$. This function is normalized to unity.

The Sudakov factors appearing in section \ref{sec:p-c} are defined as
\beq
S_{ab}(t) &=& s\left(x_1 m_B/\sqrt{2}, b_1\right) +s\left(x_3 m_B/\sqrt{2},
b_3\right) +s\left((1-x_3) m_B/\sqrt{2}, b_3\right) \non
&&-\frac{1}{\beta_1}\left[\ln\frac{\ln(t/\Lambda)}{-\ln(b_1\Lambda)}
+\ln\frac{\ln(t/\Lambda)}{-\ln(b_3\Lambda)}\right],
\label{wp}\\
S_{cd}(t) &=& s\left(x_1 m_B/\sqrt{2}, b_1\right)
 +s\left(x_2 m_B/\sqrt{2}, b_2\right)
+s\left((1-x_2) m_B/\sqrt{2}, b_2\right) \non
 && +s\left(x_3
m_B/\sqrt{2}, b_1\right) +s\left((1-x_3) m_B/\sqrt{2}, b_1\right)
\non
 & &-\frac{1}{\beta_1}\left[2
\ln\frac{\ln(t/\Lambda)}{-\ln(b_1\Lambda)}
+\ln\frac{\ln(t/\Lambda)}{-\ln(b_2\Lambda)}\right],
\label{Sc}\\
S_{ef}(t) &=& s\left(x_1 m_B/\sqrt{2}, b_1\right)
 +s\left(x_2 m_B/\sqrt{2}, b_2\right)
+s\left((1-x_2) m_B/\sqrt{2}, b_2\right) \non
 && +s\left(x_3
m_B/\sqrt{2}, b_2\right) +s\left((1-x_3) m_B/\sqrt{2}, b_2\right)
\non
 &
&-\frac{1}{\beta_1}\left[\ln\frac{\ln(t/\Lambda)}{-\ln(b_1\Lambda)}
+2\ln\frac{\ln(t/\Lambda)}{-\ln(b_2\Lambda)}\right],
\label{Se}\\
S_{gh}(t) &=& s\left(x_2 m_B/\sqrt{2}, b_2\right)
 +s\left(x_3 m_B/\sqrt{2}, b_3\right)
+s\left((1-x_2) m_B/\sqrt{2}, b_2\right) \non
 &+& s\left((1-x_3)
m_B/\sqrt{2}, b_3\right)
-\frac{1}{\beta_1}\left[\ln\frac{\ln(t/\Lambda)}{-\ln(b_2\Lambda)}
+\ln\frac{\ln(t/\Lambda)}{-\ln(b_3\Lambda)}\right], \label{ww}
\eeq
where the function $s(q,b)$ are defined in the Appendix A of
Ref.~\cite{luy01}. The scale $t_i$'s in the above equations are chosen as
\beq
t_{e}^1 &=& {\rm max}(\sqrt{x_3} m_B,1/b_1,1/b_3)\;,\non
t_{e}^2 &=& {\rm max}(\sqrt{x_1}m_B,1/b_1,1/b_3)\;,\non
t_{e}^3 &=& {\rm max}(\sqrt{x_3}m_B,1/b_2,1/b_3)\;,\non
t_{e}^4 &=& {\rm max}(\sqrt{x_2}m_B,1/b_2,1/b_3)\;,\non
t_{f} &=& {\rm max}(\sqrt{x_1 x_3}m_B, \sqrt{(x_1-x_2) x_3}m_B,1/b_1,1/b_2)\;,\non
t_{f}^3 &=& {\rm max}(\sqrt{x_1+x_2+x_3-x_1 x_3-x_2 x_3}m_B,
    \sqrt{x_2 x_3} m_B,1/b_1,1/b_2)\;,\non
t_{f}^4 &=&{\rm max}(\sqrt{x_2 x_3} m_B,\sqrt{(x_1-x_2)
x_3}m_B,1/b_1,1/b_2)\; .
\eeq
They are given as the maximum energy scale appearing in each diagram
to kill the large logarithmic radiative corrections.

\end{appendix}


\end{document}